\documentclass[a4paper]{jpconf}
\usepackage{graphicx}
\begin{document}

\title{VIP-2 at LNGS: An experiment on the validity of the Pauli Exclusion Principle for electrons}

\author{{\underline {J. Marton}}$^{1}$, S. Bartalucci$^{2}$, A. Bassi$^{3}$, M. Bazzi$^{2}$, S. Bertolucci$^{4}$, C. Berucci$^{1,2}$, M. Bragadireanu$^{2,3}$, M. Cargnelli$^{1}$, A. Clozza$^{2}$, C. Curceanu$^{2,3,9}$, L. De Paolis$^{2}$, S. Di Matteo$^{6}$, S. Donadi$^{4}$, J.-P. Egger$^{7}$, C. Guaraldo$^{2}$, M. Iliescu$^{2}$, M. Laubenstein$^{8}$, E. Milotti$^{4}$, A. Pichler$^{1}$, D. Pietreanu$^{2,3}$, K. Piscicchia$^{2,9}$, A. Scordo$^{2}$, H. Shi$^{2}$, D. Sirghi$^{2,3}$, F. Sirghi$^{2,3}$, L. Sperandio$^{2}$, O. Vazquez-Doce$^{10}$, E. Widmann$^{1}$ and J. Zmeskal$^{1}$}
\address{$^{1}$ Stefan Meyer Institute for subatomic physics, Boltzmanngasse 3, 1090 Vienna, Austria}
\address{$^{2}$ INFN, Laboratori Nazionali di Frascati, CP 13, Via E. Fermi 40, I-00044, Frascati (Roma),
Italy}
\address{$^{3}$ “Horia Hulubei”National Institute of Physics and Nuclear Engineering, Str. Atomistilor no.
407, P.O. Box MG-6, Bucharest - Magurele, Romania}
\address{$^{4}$ Dipartimento di Fisica, Universit`a di Trieste and INFN– Sezione di Trieste, Via Valerio, 2,
I-34127 Trieste, Italy}
\address{$^{5}$ University and INFN Bologna, Via Irnerio 46, I-40126, Bologna, Italy}
%The Stefan Meyer Institute for Subatomic Physics, Boltzmanngasse 3, A-1090 Vienna,
%Austria
\address{$^{6}$ Institut de Physique UMR CNRS-UR1 6251, Universit´e de Rennes1, F-35042 Rennes, France}
\address{$^{7}$ Institut de Physique, Universit´e de Neuchˆatel, 1 rue A.-L. Breguet, CH-2000 Neuchˆatel,
Switzerland}
\address{$^{8}$ INFN, Laboratori Nazionali del Gran Sasso, S.S. 17/bis, I-67010 Assergi (AQ), Italy}
\address{$^{9}$ Museo Storico della Fisica e Centro Studi e Ricerche “Enrico Fermi”, Roma, Italy}
\address{$^{10}$ Excellence Cluster Universe, Technische Universit¨at M¨unchen, Garching, Germany}

\ead{johann.marton@oeaw.ac.at}

\begin{abstract}
We are experimentally investigating possible violations of standard quantum mechanics predictions in the Gran Sasso underground laboratory in Italy. We test with high precision the Pauli Exclusion Principle and the collapse of the wave function  (collapse models). We present our method of searching for possible small violations of the Pauli Exclusion Principle (PEP) for electrons, through the search for ``anomalous'' X-ray transitions in copper atoms. These transitions are produced by ``new'' electrons (brought inside the copper bar by circulating current) which can have the possibility to undergo Pauli-forbidden transition to the 1s level already occupied by two electrons. We describe the VIP2 (VIolation of the Pauli Exclusion Principle) experimental  data taking at the Gran Sasso underground laboratories. The goal of VIP2 is to test the PEP for electrons in agreement with the Messiah-Greenberg superselection rule with unprecedented accuracy, down to a limit in the probability that PEP is violated at the level of 10$^{-31}$. We show preliminary experimental results and discuss implications of a possible violation.
\end{abstract}

\section{Introduction}
W. Pauli discovered the famous Exclusion Principle named after him which explained the periodic table of the elements \cite{Pauli1925,Nobelprize1946} (Nobel prize in 1945). The Pauli Exclusion Principle is one of the most important rules of nature and it has many consequences not only related to the periodic system of the elements but also to the stability of matter, the existence/stability of neutron stars and many other phenomena. We know that the Pauli Exclusion Principle (PEP) is itself a consequence of the spin-statistics theorem which divides nature in fermionic and bosonic systems. In spite of all efforts, no simple intuitive explanation for the PEP could be given - but several proofs of the Spin-statistics relation (Pauli exclusion principle) based on complicated arguments can be found in the literature \cite{Pauli1940,Luders1958}.
The proof by L\"uders and Zumino \cite{Luders1958} is based on a clear set of assumptions:

\begin{itemize}
  \item[-] Invariance with respect to the proper inhomogeneous Lorentz group
  \item[-] Two operators of the same field at points separated by a spacelike interval either commute or anticommute (locality)
  \item[-] The vacuum is the state of lowest energy
  \item[-] The metric of the Hilbert space is positive definite
  \item[-] The vacuum is not identically annihilated by a field
\end{itemize}

If at least one of these assumptions is invalid then a violation of the Pauli Principle would be possible.
There are also theoretical attempts to accomplish PEP violations. Some recent theoretical studies can be found in refs. \cite{Jackson2008a, Balachandran2010}.

\section{Experimental tests of the Exclusion Principle}

We know the PEP seems to be fulfilled to a high degree since no violations are found up-to-now. However, due to the outstanding importance of PEP in physics, experimental investigations were performed on many different systems: atomic transitions, nuclear transitions, nuclear reactions, anomalous atomic structure, anomalous nuclear structure, statistics of neutrinos, astrophysics and cosmology. \\

The different experimental approaches of PEP tests are based on various assumptions. According to S. Elliott \cite{Elliott2012} these experiments need to be distinguished in relation to the Messiah-Greenberg super-selection rule \cite{Messiah1964}. This rule states that the exchange symmetry of a steady state is constant in time. As a consequence, the symmetry of a quantum state can only change if a particle which is new to the system, interacts with the state.
Some experiments investigating Pauli violation in stable states resulted in remarkable upper bounds for violation  \cite{Bernabei1997, Bernabei2010, Borexino2004, Bellini2010}. However, there is the caveat that in these experimental cases the Messiah-Greenberg super-selection rule is not obeyed, meaning that one is testing an other fundamental rule, i.e. the stability of particles (e.g. electron decay \cite{Abgrall2016}).

\begin{figure}[h]
  \centering
  \includegraphics[width=12cm]{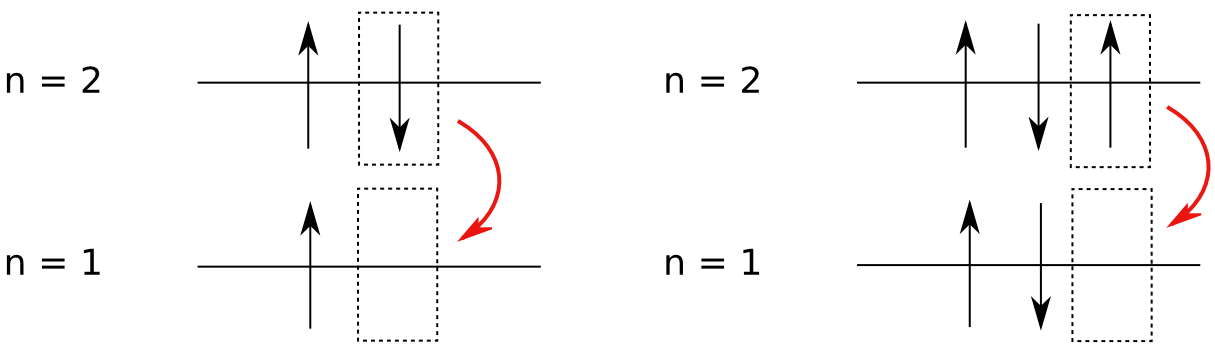}
  \caption{Transitions to the 1s ground state: Allowed transition 2p-1s (left) and Pauli forbidden transition to the fully occupied 1s state (right).}\label{scheme}
\end{figure}

A pioneering experiment was performed by Ramberg and Snow \cite{Ramberg1990} which searched for Pauli forbidden X-ray transitions in copper after introducing ``new'' electrons to the system. The concept is based on the assumption that an electric current running through a copper conductor represents a source of electrons which are ``new'' to the systems of copper atoms of the copper conductor. Thus one can search for Pauli-forbidden transitions in the copper atoms (see fig. \ref{scheme}). The transition energy of the PEP violating transition is shifted in energy due to the shielding by the ``extra'' electron in the 1s state. These shifted transition energies can be calculated using a multiconfiguration Dirac-Fock approach taking the relevant corrections (e.g. relativistic corrections) into account \cite{Sperandio2008, DiMatteo2005}.
Ramberg and Snow conducted the experiment in the basement of Fermilab and obtained the result

\begin{equation}\label{RS}
  \beta^{2}/2 \leq 1.7 \times 10^{-26}
\end{equation}

The quantity $\beta^{2}/2$ stands for the probability of a Pauli violating atomic transition and is de-facto standard in the literature.

A critical point is the question how one can identify electrons in the conductor as "new" electrons to the atoms of the conductor. Indeed the de-coherence time (at room temperature) is very short - supporting the picture of ``new'' electrons (Edoardo Milotti, private communication, 2016).

\subsection{VIP at LNGS}

\begin{figure}[h]
  \centering
  \includegraphics[width=10cm]{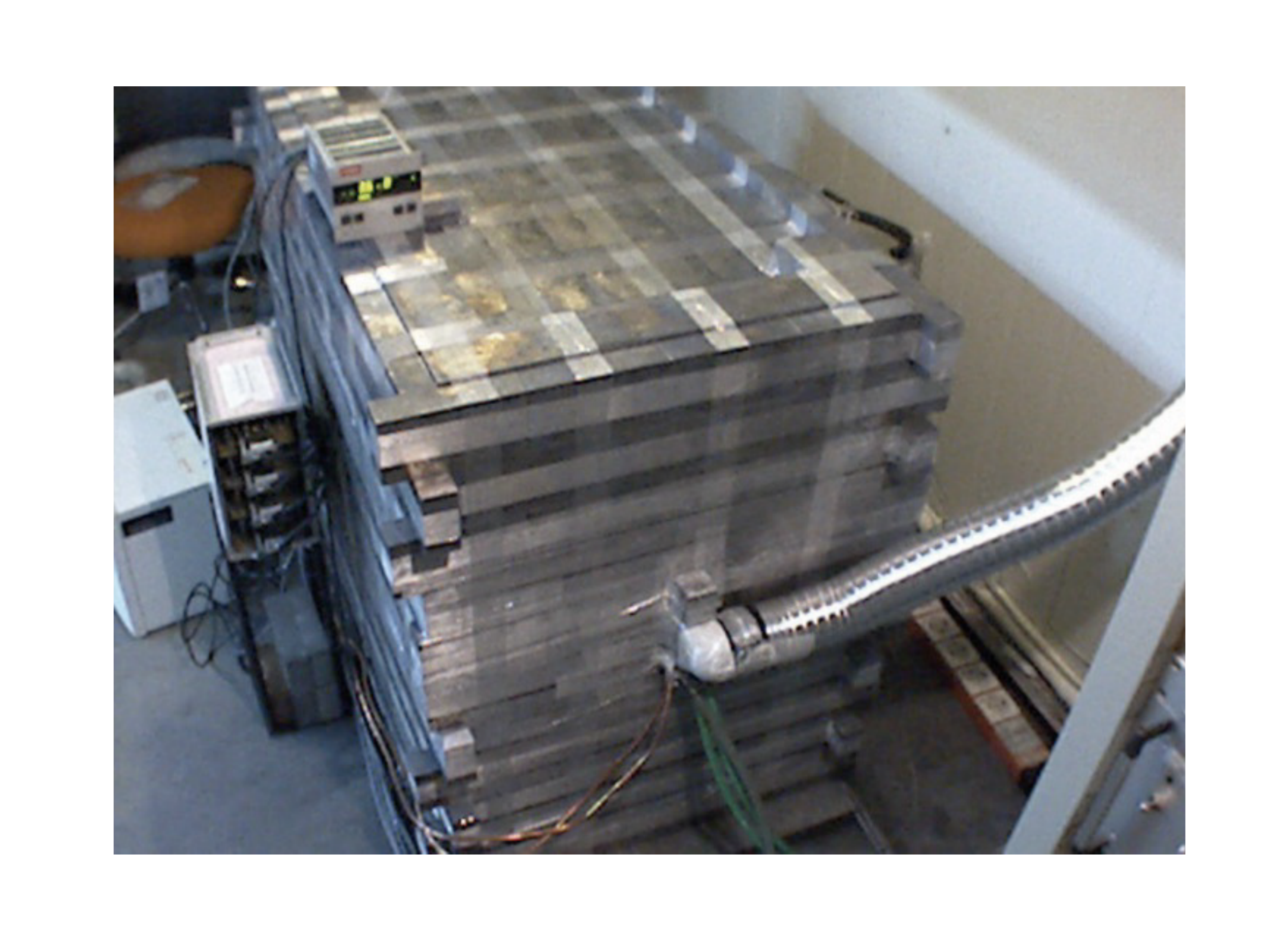}
  \caption{The VIP experiment with the lead shielding mounted in operation at LNGS.}\label{vip}
\end{figure}

A much improved experiment VIP \cite{Collaboration2004, Bartalucci2006} following the concept of Ramberg and Snow was set up in the underground laboratory LNGS in Gran Sasso/Italy (LNGS). VIP used charge coupled devices (CCDs) \cite{Egger1993} as X-ray detectors with very good energy resolution, large area and high intrinsic efficiency. The CCDs were previously successfully employed in an experiment on kaonic atoms at LNF Frascati \cite{Beer2005, Ishiwatari2006a}. The CCDs were positioned around a pure copper cylinder operated without and with up to 40 A current. The cosmic background in the LNGS laboratory is strongly suppressed ($\sim 10^{-6}$) due to the rock coverage. Additionally the setup was covered by massive lead shielding (see fig. \ref{vip}).

The VIP experiment took data alternating runs with and without current. It obtained for $\beta^{2}/2$ \cite{Curceanu2015, Piscicchia2015b}

\begin{equation}\label{beta}
  \beta^{2}/2 \leq 4.7 \times 10^{-29}
\end{equation}

Compared to the result of Ramberg-Snow it is an improvement by nearly 3 orders of magnitude.

\subsection{VIP2 at LNGS}
As a next step the experiment VIP2 with SDDs (Silicon Drift Detectors) as X-ray detectors was built and installed in LNGS. The experiment is designed for higher sensitivity by providing a larger X-ray detector solid angle, higher current and employing active shielding by plastic scintillators readout by silicon photomultipliers as background sensitive detectors. Due to the timing capability of SDDs the timing information of the SDD detectors and plastic scintillator signals can be used to additionally suppress background events.

\begin{figure}
  \centering
  \includegraphics[width=7cm]{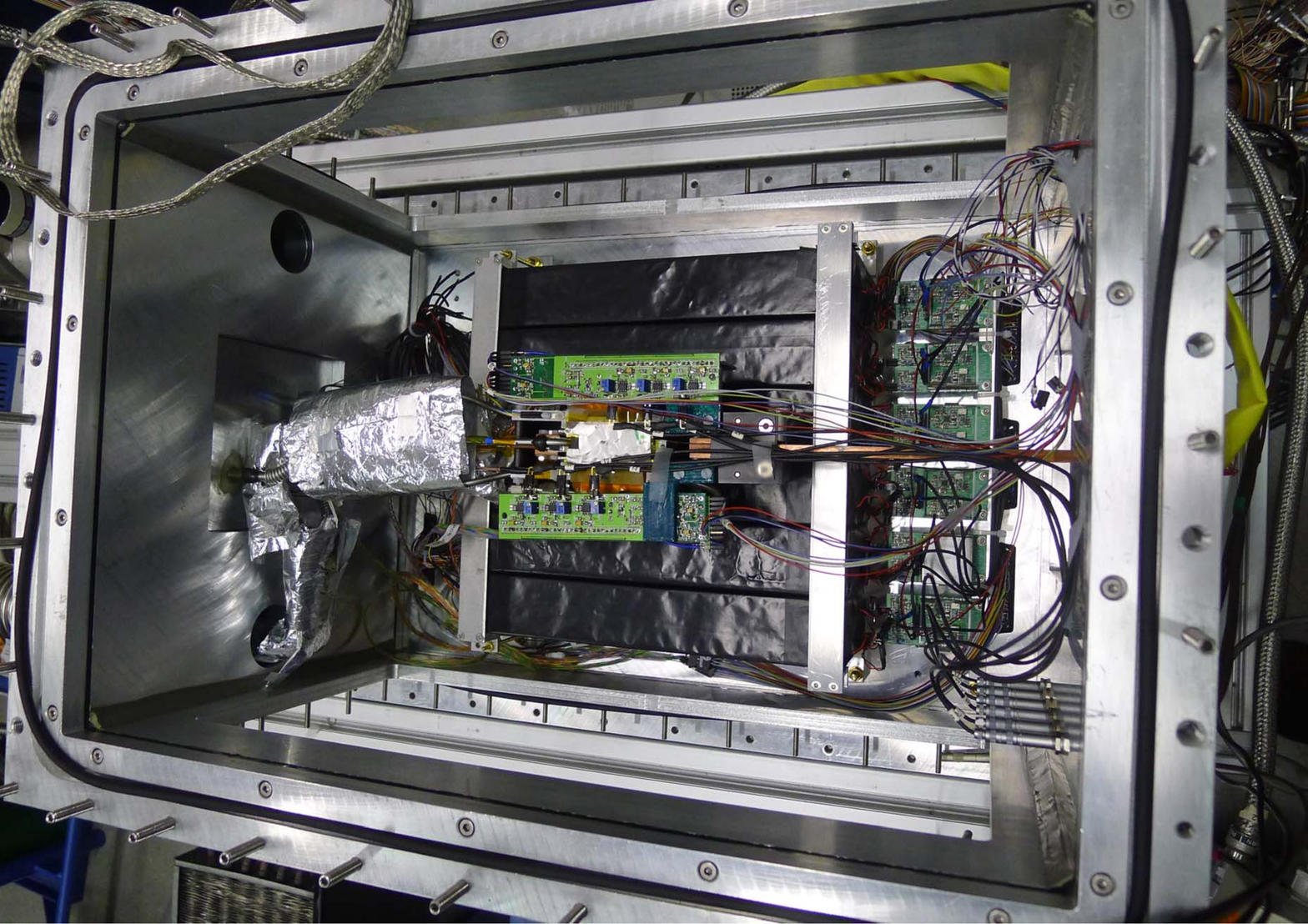}
  \caption{This photo shows the interior of the VIP2 experiment. In the box the copper target, the SDDs and the plastic scintillators are mounted. An insulation vacuum ($\sim$10$^{-5}$ mbar) inside the box is necessary to operate the SDDs at 100 K.}\label{vip2 box}
\end{figure}

\begin{figure}
  \centering
  \includegraphics[width=10cm]{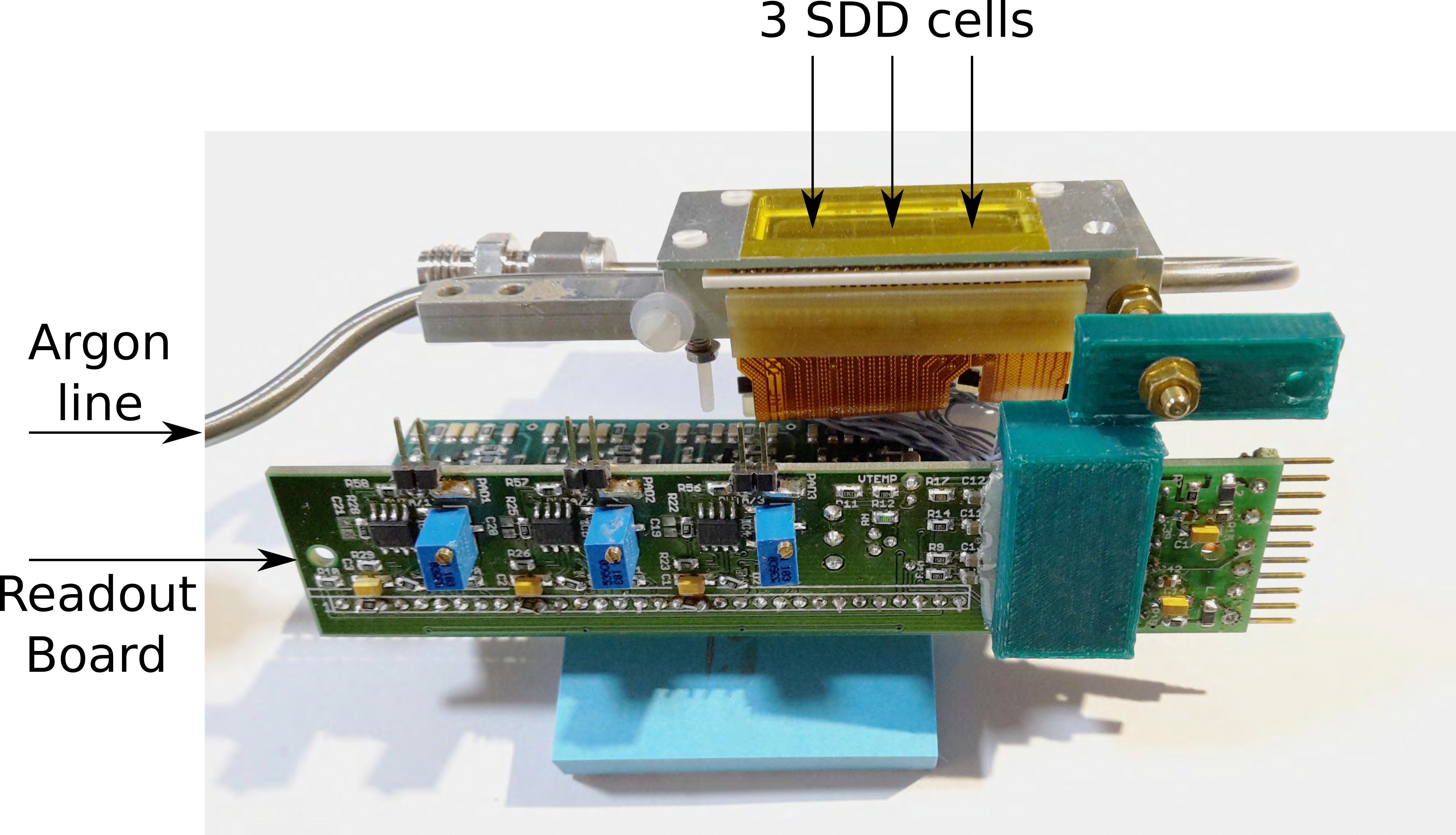}
  \caption{X-ray detector system of the VIP2 experiment. The 3 SDD cell detector is cooled by liquid argon to about 100 K and read out via the readout board.}\label{xray-de}
\end{figure}

\section{Recent Results}

\begin{figure}
  \centering
  \includegraphics[width=16cm]{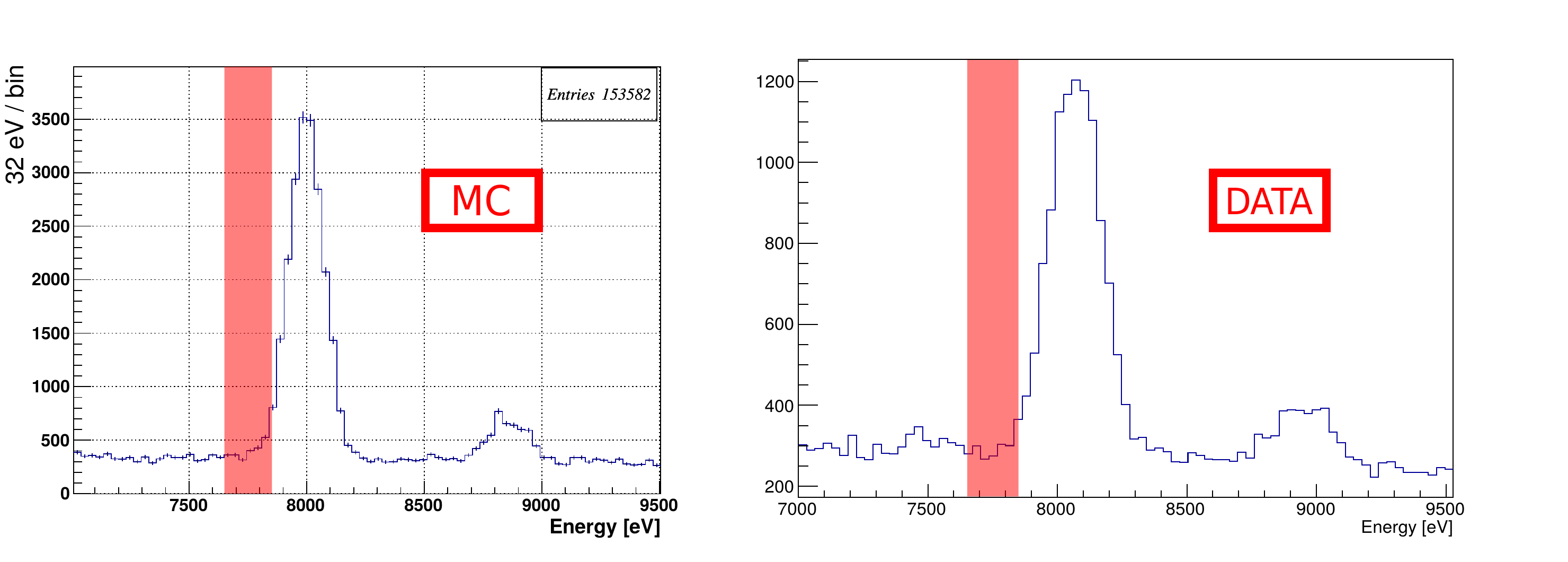}
  \caption{Comparison between 30 days of Monte Carlo simulation data (left) and 30 days of data measured at LNGS (right). The background in the region of interest (marked in red) differs only by around 30 \%.}\label{spectrum}
\end{figure}

The progress of the VIP2 experiment has  been reported in \cite{Pichler2016, Shi2016, Marton2015, Curceanu2016a}. In 2016 we collected data in a time period of $\sim$70 days without current and $\sim$40 days with 100 A current. In fig. \ref{spectrum} the Monte Carlo generated X-ray energy spectrum in the range 7-9.5 keV around the region of interest (marked in red) and the corresponding measured energy spectrum are displayed.

 In order to compare our preliminary result we used the  analysis technique of Ramberg and Snow \cite{Ramberg1990}. The analysis of this data set leads to a preliminary upper limit for the probability that the PEP is violated for electrons in copper

\begin{equation}\label{Prel-result}
  \beta^{2}/2 \leq 1.4 \times 10^{-29}
\end{equation}

It has to be emphasized that this preliminary result represents already the most stringent test of the PEP with no violation of the Messiah-Greenberg superselection rule.

\section{Summary and Outlook}

The experimental program for testing a possible PEP violation for electrons made great progress in 2016. The use of a new type of SDDs as X-ray detectors can further enhance the sensitivity by providing larger sensitive area. Furthermore, the cooling can be made more simple changing from liquid argon to Peltier cooling.
Concerning the reduction of the X-ray background we will install a passive shielding with Teflon, lead and copper. Given a running time of 3 years and alternating measurement with and without current we expect either to lower the upper limit of PEP violation by about two orders of magnitude compared to the former VIP experiment or to discover the violation.

\ack
We thank H. Schneider, L. Stohwasser, and D. St¨uckler from Stefan-Meyer-Institut for their
fundamental contribution in designing and building the VIP2 setup.
We acknowledge the very important assistance of the INFN-LNGS laboratory staff during all
phases of preparation, installation and data taking.
The support from the EU COST Action CA 15220 is gratefully acknowledged.
We thank the Austrian Science Foundation (FWF) which supports the VIP2 project with the
grants P25529-N20 and  W1252-N27 (doctoral college particles and interactions) and Centro Fermi for the grant “Problemi aperti nella meccania quantistica”.
Furthermore, this paper was made possible through the support of a grant from the
Foundational Questions Institute, FOXi (“Events” as we see them: experimental test of the
collapse models as a solution of the measurement problem) and a grant from the John Templeton
Foundation (ID 581589). The opinions expressed in this publication are those of the authors
and do not necessarily reflect the views of the John Templeton Foundation.
\\ \\ \\ \\
\bibliographystyle{iopart-num}
%
%\clearpage
\textbf{\Large References}\\ \\
\bibliography{Bibliography}{}
\end{document}